\documentclass[12pt,letterpaper]{article}
\usepackage{setspace}
\def\mboost{M_{boost}}
\def\madm{M_{ADM}}

\begin{document}

\begin{titlepage}
\vfill
\begin{flushright}
\end{flushright}
\vfill
\begin{center}
\baselineskip=16pt
{\Large\bf Boost Mass and the Mechanics of Accelerated Black Holes}
\vskip .75cm

{\large {\sl }}
\vskip 10.mm
{\bf Koushik Dutta, Sourya Ray and Jennie Traschen\footnote{email: koushik, sourya, traschen@physics.umass.edu}} \\
\vskip 0.5cm
{

    Department of Physics\\
    University of Massachusetts\\
    Amherst, MA 01003, USA\\
}
\vspace{6pt}
\today
\end{center}
\vskip .5in
\par
\begin{center}
Dedicated to Rafael Sorkin for his many contributions to our \\ understanding of physics.
\end{center}
\begin{center}
{\bf ABSTRACT}
\end{center}

\begin{quote}
In this paper we study the concept of the {\it boost mass} of a spacetime and
investigate how variations in the boost mass enter into the laws of
black hole mechanics.  We define the boost mass as the gravitational
charge associated with an asymptotic boost symmetry, similiar to how the ADM mass is
associated with an asymptotic time translation symmetry. In distinction to the ADM mass,
the boost mass is a relevant concept when the spacetime has
stress energy at infinity, and so the spacetime is not asymptotically flat.
We prove a version of the first law which relates the variation in the
boost mass to the change in the area of the black hole horizon, plus the change
in the area of an {\it acceleration horizon}, which is necessarily
present with the boost Killing field, as we discuss. The C-metric and
Ernst metric are two known analytical solutions to Einstein-Maxwell theory describing accelerating black holes which illustrate these concepts.

\vfill
\vskip 2.mm
\end{quote}
\end{titlepage}

\section{Introduction}

The close association between symmetries and conservation laws is a truth universally acknowledged amongst physicists.  Invariance under time
translation is associated with a conserved energy, or mass; spatial translations with conserved momentum; and rotational symmetry with conserved
angular momentum.  Perhaps the most famous symmetry of Minkowski spacetime is its invariance under Lorentz transformations or {\it boosts}. However,
a {\it boost charge} is conspicuously absent from the preceding list. The purpose of this paper is to study this seemingly neglected type of charge
in the context of general relativity. In general relativity, spacetime symmetries are generated by Killing vector fields.  If a spacetime with
stress-energy $T_{ab}$  has a Killing vector $V^ a$, one way to obtain a conserved charge is via the conserved current  $j^a= T^a_{\;\; b} V ^b$. The
integral of the normal component of $j^a$ over a spacelike hypersurface is conserved. However, conserved quantities also exist for spacetimes that
have  symmetries only asymptotically. In asymptotically flat spacetimes, for example, the ADM mass, momentum and angular momentum are conserved and
defined by surface integrals evaluated at spatial infinity \cite{ADM:1962}.  Abbott and Deser \cite{Abbott:1982ff} showed that similar conserved
charges, also defined as boundary integrals at infinity, exist for any class of spacetimes that are asymptotic at spatial infinity to a fixed
background metric having symmetries. For each Killing vector $V^a$ of this reference metric $ g_{ab}^{\scriptscriptstyle{(ref)}}$, there is a
conserved charge which we can denote generally as $Q\, [V^a, g_{ab}^{\scriptscriptstyle{(ref)}} ]$. Regge and Teitelboim \cite{Regge:1974zd} showed
how the original ADM charges arise as boundary terms in the Hamiltonian formulation of general relativity with asymptotically flat boundary
conditions. Hawking and Horowitz \cite{Hawking:1995fd} found similar results for the case of a general asymptotic background. They define
gravitational charges by varying the Einstein action.

It is clear then that for asymptotically flat spacetimes, there exists a conserved charge, corresponding to the boost symmetry of Minkowski
spacetime, which we will call $M_{boost}$.  It is also relatively simple to understand why the boost mass $M_{boost}$ is neglected in most
considerations of asymptotically flat spacetimes. The ADM mass, $M_{ADM}$, and the boost mass, $M_{boost}$, are, of course, zero for Minkowski
spacetime. One can show that if $M_{ADM}$ is increased from zero, then $M_{boost}$ becomes infinite.  This happens essentially because the boost
Killing vectors of Minkowski spacetime diverge at spatial infinity, while the time translation Killing vector is constant.  According to the positive
energy theorem $M_{ADM}$ vanishes only in  Minkowski spacetime \cite{Witten:1981mf}.  Therefore,  once the metric deviates from Minkowski spacetime,
the boost mass necessarily  becomes infinite. This is discussed in more detail in section \ref{mboost}.  We then see that $M_{boost}$ is
uninteresting for asymptotically flat spacetimes and that to examine its properties, we need to find a more appropriate physical setting
\footnote{References \cite{Unruch and Wald, Hawking:1994ii} define the boost energy as the volume integral of the time component of the local current
$\int dv( n_a j^a )$. This boost energy is not the same as the boundary term that serves as our definition of $M_{boost}$. This is discussed in
section \ref{def}.}.

If the background spacetime contains stress-energy in the asymptotic region, then the story is different in an interesting way.  For example, suppose
that the background spacetime is a static, straight cosmic string. This is not asymptotically flat at spatial infinity, because the string extends to
infinity\footnote{It is asymptotically locally flat if one moves to spatial infinity only in directions transverse to the string}. We will model the
cosmic string metric by flat spacetime minus a wedge, and denote the resulting metric by  $\eta^{\scriptscriptstyle{(\nu)}}_{ab}$, where $\nu$
parameterizes the missing angle. The metric $\eta^{\scriptscriptstyle{(\nu)}}_{ab}$ is symmetric with respect to a time translation Killing vector
$T^a$, with respect to translations along the string and rotations around it, and also with respect to boosts along the string generated by the
Killing vector $\xi^a$.   In this paper, we will define gravitational charges for spacetimes that are asymptotic to
$\eta^{\scriptscriptstyle{(\nu)}}_{ab}$.  Of course, if the infinite string itself is compared with Minkowski spacetime, then it will have an
infinite mass. However, the charges we compute represent the finite residual contributions after subtracting off the contribution of the infinite
string.  The development parallels the asymptotically flat case. If the metric is everywhere exactly $\eta^{\scriptscriptstyle{(\nu)}}_{ab}$, then
the charges $Q[V^a ,\eta^{\scriptscriptstyle{(\nu)}}_{ab} ]$ vanish\footnote{If the reference metric is clear from the context, we will simply write
$Q[V^a]$, rather than {\it e.g.} $Q[V^a ;\eta^{\scriptscriptstyle{(\nu)}}_{ab} ]$ in the following. }. Also as in the asymptotically flat case, we
will continue to call the charge $Q[T^a]$ the ADM mass, and $Q[\xi^a ]$ the boost mass. However, it is now possible to find spacetimes that have
$M_{ADM}=0$, but are not everywhere equal to $\eta^{\scriptscriptstyle{(\nu)}}_{ab}$!

For example, imagine clipping a segment of the string from the interior, and adding a ball of mass of some sort to each free end.  If this is done so
that, roughly speaking, we are just moving mass around, then $M_{ADM}$ remains zero, but $M_{boost}$ becomes nonzero. The C-metric provides a well
known example of such a rearrangement, in which the balls of mass at the string ends are themselves black holes \cite{C,bicak}.  As a second example,
let the background reference metric be the Melvin spacetime, which has a non-vanishing magnetic field everywhere \cite {melvin}. Again, this is not
asymptotically flat, and because the background has stress-energy, we can again imagine rearranging the mass in the interior, such that the monopole
moment of the mass distribution is unchanged in the far field.  Since $M_{ADM}$ measures this monopole moment, it would still vanish. On the other
hand, the boost mass $M_{boost}$, we will see, essentially measures the dipole moment of the mass distribution and would be both non-infinite and
generically nonzero. The Ernst spacetime is an example of this sort of rearrangement, in which two charged black holes are accelerated apart by the
background magnetic field \cite{Ernst}. We will discuss these two examples, the C-metric and the Ernst metric, below in sections \ref{cst} and
\ref{Ernstst} respectively.

The reader may wonder how we distinguish  between a {\it boost} Killing vector and a {\it time translation} Killing vector  in a general spacetime.
We will call $\xi^{a}$ a boost Killing vector if it is timelike in a region, which includes a part of infinity, and which is bounded in the interior
by an acceleration horizon.  By an acceleration horizon we will mean a surface where $\xi^{a}$ becomes null, and which has noncompact spatial slices.
These definitions are motivated by the behavior of the boost Killing vector $\xi^a = x({\partial\over\partial t})^a +t ({\partial\over\partial x})^a$
in Minkowski spacetime and are discussed in more detail in section \ref{definitions}.

After studying $M_{boost}$ in the context of these examples, we will turn our attention to generalizations of the first law of black hole mechanics,
or thermodynamics, that take into account variations, $\delta\mboost$, in the boost mass. The usual first law of black hole thermodynamics applies to
black holes in asymptotically flat spacetimes, for which the generator of the black hole horizon is a Killing vector $T^a$ that becomes time
translation at infinity.  The Killing vector $T^a$ is timelike outside the black hole and becomes null on the horizon. The  first law links
variations of properties of the spacetime evaluated at infinity to variations evaluated at the black hole horizon.  For example, in the simplest
uncharged and non-rotating case, we have
\begin{equation}
\delta M_{ADM} ={\kappa _{\scriptscriptstyle {bh}}\over 8\pi}\; \delta A_{bh}
\end{equation}
where $\madm$  is evaluated at infinity and the horizon area $A_{bh}$ is evaluated at the horizon. Now suppose instead that the generator of the
black hole horizon is a Killing vector $\xi ^a$, which is null on the horizon, timelike in a region outside the black hole and asymptotes to a boost
at infinity.  When we say that $\xi^{a}$ is a boost we will mean that there is also an acceleration horizon, on which the Killing vector $\xi^{a}$ is
null. An analogue of the usual first law, the variations about this black hole spacetime will involve $\delta\mboost$ at infinity, $\delta A_{bh}$ at
black hole horizon, and also $\delta A_{acc}$, the variation of the area of the acceleration horizon. The contribution from the acceleration horizon
arises because it is an additional boundary on which the Killing field becomes null. In section \ref{Firstlaw}, we derive such a first law for
variations in the boost mass
\begin{equation}
\delta M_{boost} = \frac{1}{8\pi}\kappa_{\scriptscriptstyle{bh}}
\delta A_{bh} + \frac{1}{8\pi}\kappa_{\scriptscriptstyle{acc}}
\delta A_{acc}.
\end{equation}
This equation is written for purely gravitational case, in which there are no stress-energy sources and no variation in the electric charge. Such
additional terms are included in section \ref{Firstlaw}. A similar expression is discussed by Jacobson and Parentani \cite{Jacobson}. There are differences
between this earlier paper and this one and we will discuss this issue in section \ref{Firstlaw}.

In section \ref{mboost} we give the expression for $M_{boost}$ when the spacetime is  asymptotically  Rindler spacetime  minus a wedge
$\eta^{\scriptscriptstyle{(\nu)}}_{ab}$, with a specified missing angle and an acceleration parameter. We identify the fall-off conditions on the metric
such that $M_{boost}$ is finite. Later, a simple example of the theorem is provided by working out $\delta M_{boost}$ when the variation is with
respect to a nearby C-metric. The asymptotically Melvin case is considered in section \ref{Ernstst}.

\section{Perturbative Constraints on Charges}
\label{perturbativeconstraints}

In this section we show that gravitational charges can be defined in a useful way for spacetimes which approach a reference metric, whenever the
reference spacetime has a Killing field, not just in the asymptotically flat case. The definition of the charge that we present is useful, because we can then
prove a relation analogous to the usual first law of black holes for variations in the charge. Readers familiar with the result of equation
(\ref{perturbativecharge}) at the end of this section, may skip this section without loss of continuity.

\subsection{Basic Hamiltonian Formalism}

In this subsection we set up the Hamiltonian formalism of General Relativity. More details of the Hamiltonian formalism are given in references
\cite{Wald,Traschen}.

The calculations are little involved but the idea is simple. Let the spacetime $(M,g_{ab}$) be foliated by a family of spacelike slices
($\Sigma_{t}$) with a timelike vector field $\frac{\partial}{\partial t}$, and a unit normal field $n_{a} = -N\nabla_{a}t$. Let $g_{ab}$ be a
Lorentzian metric satisfying Einstein's Equation $G_{ab}= 8{\pi} T_{ab}$, and $\nabla_{a}$ be the derivative operator compatiable with $g_{ab}$,
i.e, $\nabla_{c} g_{ab} = 0$.

The spacetime metric $g_{ab}$ induces a spatial metric $s_{ab}$ on the constant time spacelike hypersurfaces $\Sigma_{t}$,
\begin{equation}
g_{ab} = s_{ab} -
n_{a}n_{b}\hspace*{5pt}, \hspace*{20pt}
n^{a}s_{ab} = 0,
\end{equation}
and $n \cdot n=-1$.

Here we will consider Einstein-Maxwell theory. The formalism can be generalized for any energy momentum tensor $T^{ab}$, describing a matter field
which has a well defined Hamiltonian formalism. In Einstein-Maxwell theory, a point in the phase space is specified by the initial data
$(s_{ab},\pi^{ab},\tilde{A}_{a}, E^{a})$ on a spacelike surface $\Sigma$, where $\tilde{A}_{a} = s_{a}^{c}A_{c}$ is the projection onto $\Sigma$ of
the spacetime gauge potential $A_{a}$. $\pi^{ab}$ is the momentum conjugate to $s_{ab}$, and is related to the extrinsic curvature $K_{ab}$ of
$\Sigma$
\begin{equation}
\pi^{ab} = \sqrt{s}(K^{ab} - s^{ab}K),
\end{equation}
where $s = det[s_{ab}]$. The momentum conjugate to $\tilde{A}_{a}$ is proportional to the electric field $E^{a}$.

Initial data must satisfy the Einstein constraints, which are non-dynamical equations. In the Hamiltonian variables, the constraints on $\Sigma$ are
\begin{equation}
0 = C = \frac{1}{4\pi}\sqrt{s}D_{a}E^{a},
\end{equation}
\begin{equation}
0 = C_{0} = \frac{1}{16\pi}\sqrt{s}[- R] + 2E_{a}E^{a} +
\tilde{F}_{ab}\tilde{F}^{ab} + \frac{1}{s}(\pi^{ab}\pi_{ab} - \frac{1}{2}\pi^{2}),
\end{equation}
\begin{equation}
0 = C_{a} = -\frac{1}{8\pi}\sqrt{s}[D_{b}(\pi^{b}_{a}/\sqrt{s}) -
2\tilde{F}_{ab}E^{b}],
\end{equation}
where $D_{a}$ is the derivative operator on $\Sigma$ compatible with $s_{ab}$, $R$ denotes the scalar curvature of $s_{ab}$ and
$\tilde{F}_{ab}=2D_{[a}\tilde{A}_{b]}$. The Hamiltonian ${\cal{H}}_{tot}$ for Einstein-Maxwell theory is a sum of the constraints,
\begin{equation}
{\cal{H}}_{tot} = NC_{0} + N^{a}C_{a}+A_{t}C\equiv(N,N^{a},A_{t})\cdot H_{tot}
\end{equation}
Here $N$, $N^{a}$ and $A_{t}$ are Lagrange multipliers, which may be prescribed arbitrarily. The variations of the Hamiltonian with respect to the
Lagrange multipliers give the constraint equations, and the usual Hamilton's equations give the evolution of the dynamical variables
$(s_{ab},\pi^{ab},\tilde{A}_{a},E^{a})$. The vector $w^{a}=Nn^{a}+N^{a}$ represents the flow of time in the spacetime and the Hamiltonian generates
this time flow. The projection of $\vec{w}$ onto $\Sigma$ yields the shift vector $N^{a}$ and the projection normal to $\Sigma$ yields the lapse
function $N$. ${\cal{H}}_{tot}$ is identically zero on solutions.

\subsection{Definition of Gravitational Charges}\label{def}

Let $g_{ab}^{\scriptscriptstyle{(ref)}}$ be a fixed spacetime which we call the reference, and which has a Killing field $V^{a}$. Let $g_{ab}$ be a
metric which asymptotes to the reference spacetime $g^{\scriptscriptstyle(ref)}_{ab}$. Let $ \ ^{\scriptscriptstyle{(4)}}\gamma_{ab} =g_{ab} -
g_{ab}^{\scriptscriptstyle(ref)}$ be the difference between the spacetime metric and the reference metric. Note that $ \ ^{\scriptscriptstyle{(4)}}\gamma_{ab}$ is the perturbation to the full spacetime metric, distinguished from $\gamma_{ab}$, the
perturbation to the spatial metric. The definition of $Q_V$, the gravitational charge
associated with the asymptotic Killing field $V^a$, will depend on $\gamma_{ab}$  only in the asymptotic region, on
$g^{\scriptscriptstyle(ref)}_{ab}$, and on the boundary $\partial\Sigma_{asy}$ of the volume $\Sigma$. The idea is that we define $Q_{V}$ as a
boundary integral of the form
\begin{equation}
Q_V( g^{\scriptscriptstyle(ref)} , g) =
\frac{1}{16\pi}\int_{\partial \Sigma_{asy}} B^{c} [s^{\scriptscriptstyle(ref)},
 \pi ^{\scriptscriptstyle(ref)}, \gamma ,\delta \pi , V]
da_{c}, \label{charge}
\end{equation}
and we will choose the integrand $B^{c}$ such that there is a theorem of the form of the first law for the variations of $Q_{V}$. The expression for the
boundary integrand $B^c$ is given in equations (\ref{gravboun}) and (\ref{maxboun}). We will often shorten the notation to $Q_V$.

So far we have not said anything about the rate at which $ \ ^{\scriptscriptstyle{(4)}}\gamma_{ab}$ must go to zero. Indeed, for a particular metric
$g_{ab}$ and hence a particular $ \ ^{\scriptscriptstyle{(4)}}\gamma_{ab}$, the charge might be finite for one Killing vector, but infinite or zero
for a different Killing vector. One of the interesting issues in  sorting out the meaning of the boost mass, is to understand when it is infinite, or
finite, or zero. We do this in section \ref{mboost}. Depending on the background and on the spacetime of interest, either the boost mass or the ADM mass will give useful information,
but not both.

If the reference metric is  Minkowski, then asymptotically the spacetime has all the Poincare symmetries, and we can define Killing charges
corresponding to all generators of the  Poincare group. The $Q_V$ are the different conserved charges, as Killing vector $V^a$ ranges over all the
generators. The integral in equation (\ref{charge}) reduce to the usual ADM mass and angular momentum when the Killing vector is taken to be time
translation or a rotation respectively. Regge and Teitleboim \cite{Regge:1974zd} have shown that the boundary terms satisfy the correct algebra of
Minkowski spacetime.

For two perturbatively close metrics $g_{ab}$ and $g_{ab}^{\scriptscriptstyle(0)}$, we will next use this definition to prove a theorem about the
variations of the Killing charge $\delta Q $ which is defined by the same expression as in equation (\ref{charge}), with
$g_{ab}=g_{ab}^{\scriptscriptstyle(0)}+ \lambda \ ^{\scriptscriptstyle{(4)}}h_{ab} + \mathcal{O}(\lambda^{2})$, $ \ ^{\scriptscriptstyle{(4)}}h_{ab}$ now being
the perturbation to the metric $g_{ab}^{\scriptscriptstyle(0)}$. Both $g_{ab}$ and $g_{ab}^{\scriptscriptstyle(0)}$ asymptote to the
metric $g_{ab}^{\scriptscriptstyle(ref)}$. In this case we consider $g_{ab}^{\scriptscriptstyle(0)}$ as our background.

We will see that $M_{boost}$, the Killing charge corresponding to the asymptotic boost symmetry of the background spacetime, will play an important role in this
paper. As previously mentioned, references \cite{Unruch and Wald, Hawking:1994ii} define a boost energy  $E_{boost}$, in spacetimes which have a
boost Killing vector everywhere. $E_{boost}$ is the integral over the volume $\Sigma$ of $n_a \xi ^b T^a _b$, where $\xi ^a$ is the boost Killing
vector. When the background is de Sitter, anti-de Sitter or Minkowski spacetime, the construction of Abbott and Deser \cite{Abbott:1982ff} shows that
the boundary term defining the boost charge $Q_{V}$ in equation (\ref{charge}) is equal to $E_{boost}$, $plus$ a volume integral of  nonlinear terms from
the Einstein tensor \footnote{More precisely, the Einstein tensor is formally expanded in $\ ^{\scriptscriptstyle{(4)}}\gamma _{ab}$, and let
$G_{ab}^{(NL)} =G_{ab}-G_{ab}^{(L)}$, where $G_{ab}^{(L)}$ is the term linear in $\gamma$. Then $Q_{boost} = E_{boost}+\int _{\Sigma}
G_{ab}^{(NL)}n^a \xi ^b$.}. This construction would have to be repeated with $g_{ab}^{\scriptscriptstyle{(0)}}$ taken to be a cosmic string or
magnetic field background, to see the same type of relation holds here.

\subsection{Gauss's Law for Perturbations}\label{gauss}

In this subsection we study solutions to the linearized Einstein equations. Let $g_{ab}$ and $g_{ab}^{\scriptscriptstyle(0)}$ both approach the
reference $g_{ab}^{\scriptscriptstyle(ref)}$, and consider the case when the two metrics are perturbatively close everywhere. Suppose that
$g^{\scriptscriptstyle{(0)}}_{ab}$ has a Killing field $V^a$. $V^a$ is a Killing field throughout the spacetime, not just asymptotically. Then
perturbations about the zeroth-order spacetme satisfy a Gauss's Law type constraint \cite{Traschen}. First we summarize the derivation of this
result. Then in section \ref{1stLaw} we outline how this result yields the first law of black hole mechanics, when
$g_{ab}^{\scriptscriptstyle{(0)}}$ is asymptotically flat and $V^a$ is a time translation Killing vector of a static black hole spacetime. Finally in
section \ref{Firstlaw} we apply the constraint to derive the first law of black hole mechanics when $g_{ab}^{\scriptscriptstyle{(0)}}$ is
asymptotic to flat spacetime minus a wedge $\eta^{\scriptscriptstyle{(\nu)}}_{ab}$, and $g_{ab}^{\scriptscriptstyle{(0)}}$ has a boost Killing vector
$\xi^a$ .

Let $(s_{ab}{\scriptstyle{(\lambda)}},\pi^{ab}{\scriptstyle{(\lambda)}},\tilde{A}_{a}{\scriptstyle{(\lambda)}},E^{a}{\scriptstyle{(\lambda)}})$ be a
one parameter family of solutions to the Einstein-Maxwell theory with pertubative expansion $ s_{ab} = s_{ab}^{\scriptscriptstyle(0)} + \lambda
h_{ab} + \mathcal{O}(\lambda^{2}) + ..,$ and $\tilde{A}_{a} = \tilde{A}_{a}^{\scriptscriptstyle(0)} + \lambda
\tilde{A}_{a}^{\scriptscriptstyle(1)} + \mathcal{O}(\lambda^{2}) + ..$, and similarly for the corresponding conjugate
momenta. $s_{ab}^{\scriptscriptstyle(0)}$ is the spatial metric induced by $g_{ab}^{\scriptscriptstyle(0)}$ on the constant time hypersurface
$\Sigma$. So, the set $(p^{\scriptscriptstyle(0)},q^{\scriptscriptstyle(0)})$ is a solution to the Einstein-Maxwell equation with a Killing vector;
$(p^{\scriptscriptstyle(1)},q^{\scriptscriptstyle(1)})$ solve the equations linearized about the zeroth order solutions, and so on. In particular,
$(p^{\scriptscriptstyle(1)},q^{\scriptscriptstyle(1)})$ solve the linearized constraints $\delta {\cal{H}}_{tot} = 0$.

Let $F,\beta^{a}$ be an arbitrary function and vector on $\Sigma$, and consider the linear combination of the constraints $(F,\beta^{a},A_{t}) \cdot
H_{tot}(s_{ab},\pi^{ab},\tilde{A}_{a},E^{a})=0$. Then the perturbative fields $(p^{\scriptscriptstyle(1)},q^{\scriptscriptstyle(1)})$ are solutions
to the following linearized constraints,
\begin{equation}
(F,\beta^{a},A_{t})\cdot \delta H_{tot} \cdot (p^{\scriptscriptstyle(1)},q^{\scriptscriptstyle(1)}) = 0.\label{linear}
\end{equation}
We can rewrite this equation in terms of the adjoint operator and
a total derivative,
\begin{equation}
(p^{\scriptscriptstyle(1)},q^{\scriptscriptstyle(1)}) \cdot \delta H^{*}_{tot} \cdot (F,\beta^{a},A_{t}) + D_{a}B^{a} = 0, \label{pertu}
\end{equation}
where $B^{a}$ is a function of the $(p^{\scriptscriptstyle(1)},q^{\scriptscriptstyle(1)})$, the Lagrange multipliers, and of course the background
spacetime. The boundary term vector $B^{a}$ is the sum of a gravitational piece and a contribution from the matter fields, $B^{a}
[s^{\scriptscriptstyle(ref)},  \pi ^{\scriptscriptstyle(ref)}, h ,\delta \pi , V]
 = B_{G}^{a} + B_{M}^{a}$, where
 \begin{equation}
B_{G}^{a} = F(D^{a}h - D_{b}h^{ab}) - hD^{a}F + h^{ab}D_{b}F
 +\frac{\beta^{b}}{\sqrt{s}}(\pi^{cd}h_{cd}s^{a}_{b} - 2\pi^{ac}h_{bc} - 2\delta
 \pi^{a}_{b}),\label{gravboun}
 \end{equation}
\begin{equation}
 B_{M}^{a} = -\frac{1}{\sqrt{s}}A_{t}\delta p^{a} +
 4F\tilde{F}^{ab}\delta \tilde{A}_{b} + \frac{2}{\sqrt{s}}\beta^{[a}p^{b]}\delta \tilde{A}_{b}.\label{maxboun}
\end{equation}
Here $h = h_{ab}s^{ab}$ and $p^{a} = -4N \sqrt{s} F^{ta}$ is the electromagnetic momentum conjugate to ${A}_{a}$.

Now, Hamilton's equations for the background spacetime are
\begin{equation}
(\dot{s}^{\scriptscriptstyle{(0)}},-\dot{\pi}^{\scriptscriptstyle{(0)}},\dot{A}^{\scriptscriptstyle{(0)}},-\dot{E}^{\scriptscriptstyle{(0)}}) =
\delta H^{*}_{tot}\cdot(F,\beta^{a},A_{t}),
\end{equation}
where $\dot{f}$ is the lie derivative of $f$ along a vector field $V^{a}=Fn^{a}+ \beta^{a}$.

Thus if $V^{a} $ is a Killing field of the background spacetime, the Lie derivatives vanish and therefore $F$ and $\beta^{a}$ are solutions to the
differential equation $\delta H^{*}_{tot} \cdot (F,\beta^{a}, A_{t}) = 0$. Equation (\ref{pertu}) then implies that all perturbations about the
background spacetime $g_{ab}^{\scriptscriptstyle{(0)}}$ must satisfy the source free Gauss's Law type constraint
\begin{equation}
D_{a}B^{a} = 0.\label{GaussLawDiff}
\end{equation}

So far we have been discussing pure Einstein-Maxwell theory. But, it is simple to include  additional perturbative sources $\delta T^{ab}$.
Then the identity modifies to $D_{a}B^{a} = \delta S$, where $\delta S = -16\pi V^{a}n^{b}\delta T_{ab}$. Using Stoke's law over the volume
$\Sigma$, we get the following integral form of the constraint,
\begin{equation}
-\int_{\Sigma}{\sqrt{s} \delta S}  = \int_{\partial
\Sigma}{da_{c}B^{c}}, \label{constraint}
\end{equation}
where $\partial \Sigma$ are all the boundaries of the volume $\Sigma$. Equations (\ref{GaussLawDiff}) and (\ref{constraint}) are the main results of
this section and key ingredients to the construction of the first law of black hole mechanics in different asymptotic backgrounds with different Killing
fields.

Analogous to equation (\ref{charge}), we define the perturbative charge $\delta Q_{V}$. Equation (\ref{constraint}) can then be rewritten as
\begin{equation}
\delta Q_{V}(g, g^{\scriptscriptstyle{(0)}})\equiv \frac{1}{16\pi}\int_{\delta \Sigma_{\infty}}da_{c}B^{c}= -\frac{1}{16\pi}\sum_{i}\int_{\delta
\Sigma_{i}}da_{c}B^{c} - \frac{1}{16\pi}\int_{\Sigma}{\sqrt{s} \delta S}\label{perturbativecharge}
\end{equation}
where the sum is over all boundaries other than the one at infinity. We emphasize that both $Q_{V}$ and $\delta Q_{V}$ depend on $g^{\scriptscriptstyle{(ref)}}_{ab}$, since both
$g_{ab}$ and $g_{ab}^{\scriptscriptstyle{(0)}}$ asymptotes to $g^{\scriptscriptstyle{(ref)}}_{ab}.$

\section{ Usual First Law of Black Hole Mechanics}
\label{1stLaw}

Readers familiar with the derivation of the first law of black hole mechanics can skip this section without loss of continuity.

As a pedagogical application of the concepts discussed in the last section, consider the case of an asymptotically flat, stationary axisymmetric
black hole spacetime satisfying Einstein's equation. We assume that the black hole event horizon is a bifurcate Killing horizon with the bifurcation
surface $\delta \Sigma_{b}$. Let $t^{\mu}$ and $\phi^{\mu}$ denote the Killing fields on this spacetime which asymptotically approach time
translation and rotation at spatial infinity respectively. So the volume $\Sigma$ is bounded by the internal compact boundary $\delta \Sigma_{b}$ and
spatial infinity. In this case there exists a linear combination $V^{a} = t^{a} + \Omega \phi^{a}$, which is the generator of the horizon, and defines
the angular velocity $\Omega$ of the horizon. $V^{a}$ vanishes on the bifurcation surface.

Using the asymptotically flat boundary conditions for the perturbations to this spacetime, the boundary terms at spatial infinity simplify.
Substituting $t^{\mu}$ and $\phi^{\mu}$ into equation (\ref{perturbativecharge}) and using $F \rightarrow 1$ at spatial infinity one finds the
standard expressions for change in mass and angular momentum respectively,
\begin{equation}
16\pi\delta M_{ADM} = \int_{\partial \Sigma_{\infty}}da_{c}(-D^{c}h +
D_{b}h^{cb}),\label{adm}
\end{equation}
\begin{equation}
16\pi\Omega \delta J = -\Omega \int_{\partial \Sigma_{\infty}}\label{deltamadm}
da_{c}\frac{2\phi^{b}\delta \pi_{b}^{c}}{\sqrt{s}}.
\end{equation}

On the bifurcation surface of the horizon $V^{a}$ vanishes, and the gravitational boundary term becomes
\begin{equation}
\int_{\partial \Sigma_{\scriptscriptstyle{bh}}}da_{c}(-hD^{c}F +
h^{cb}D_{b}F) = 2\kappa_{\scriptscriptstyle{bh}} \delta
A_{\scriptscriptstyle{bh}},
\end{equation}
where $\kappa_{\scriptscriptstyle{bh}}$ and $A_{\scriptscriptstyle{bh}}$ are the surface gravity \footnote[1]{Surface gravity $\kappa$ is defined by
the equation $V^{b}\nabla_{b}V_{a} = \kappa V_{a}$, where $V^{a}$ is the generator of the horizon} and the area of the black hole horizon
respectively. $A_{t}= V^{b}A_{b}$ vanishes on the bifurcation surface, and $A_{t} \rightarrow 1$ at spatial infinity. Assembling the boundary terms
into equation (\ref{perturbativecharge}) we have that for any asymptotically flat solutions to the linearized equations
\begin{equation}
\delta M_{ADM} = \Omega \delta J +
\frac{1}{8\pi}\kappa_{\scriptscriptstyle{bh}} \delta
A_{\scriptscriptstyle{bh}} - A_{t}\delta Q + \int_{\Sigma}V^{a}n_{b}\delta T^{b}_{a}.\label{firstlawadm}
\end{equation}
The above equation is the standard form of the first law of black hole mechanics \cite{Bardeen,Sudarsky}. It will be of interest to compare the
source term for $\delta M_{ADM}$  to the source term for $\delta M_{boost}$.

\section{Boost Mass and First Law for Black Hole} \label{boostmassand1stlaw}

In the previous section we outlined the derivation of the first Law of black hole mechanics in an asymptotically flat spacetime. This involved using
the Gauss's Law for perturbations in the integral form. The boundary terms are different if the spacetime is not asymptotically flat, and/or has
additional internal boundaries. In turn, this mean that the boundary terms may have different physical interpretations than in the asymptotically
flat case. We will now address thsese issues when the Killing field is boost $\xi^{a}$.

\subsection{Boost Killing Vectors and Acceleration Horizons}\label{definitions}

Let us turn to the definition of  a boost Killing vector and an acceleration horizon. First consider Minkowski spacetime, $ds^2 =-{dt^{\prime}} ^2+
{dx^\prime} ^2 +{dy^\prime} ^2 +{dz^\prime} ^2$. $T^a ={\partial\over \partial {t^\prime}}$ is a time translation Killing vector, and is timelike
everywhere. The boost Killing vector $\xi^a = {z^\prime}{\partial\over \partial {t^\prime}} +{t^\prime} {\partial\over \partial {z^\prime} }$ is
timelike in the wedges ${z^\prime}^2 >{t^\prime}^2$ with $ {z^\prime}>0$, and ${z^\prime}^2 >{t^\prime}^2$ with ${z^\prime}<0$. $\xi^{a}$ is null on
the surfaces ${z^\prime} =\pm {t^\prime}$. Pick one wedge, say $z^{\prime}>0$. Then the region in which $\xi ^a$ is timelike is bounded by infinity
and two null surfaces which intersect at ${z^\prime}=0$ and extend to null infinity. We will use the term ``{\it acceleration horizon}'',
${\cal{H}}_{acc}$ of $\xi^a$ to refer to this null boundary of one connected region in which $\xi^a$ is timelike. A key feature of ${\cal{H}}_{acc}$
is that its spatial sections are noncompact. For example, at ${t^\prime}=0$, ${\cal{H}}_{acc}$ is the plane ${z^\prime}=0$ which extends to spacelike
infinity. In Rindler spacetime, which is just Minkowski spacetime written in the coordinates of an observer who undergoes constant acceleration, this
is usually called the Rindler horizon. We use the term ``acceleration horizon'' instead, since it is commonly used when analyzing the C- and Ernst
metrics.

Motivated by the example of flat spacetime, and by the behavior of the analogous Killing fields in the C-metric and the Ernst metric, we will call a
Killing field a ``{\it boost}'' if it is time-like in some region  of the spacetime which is bounded by a part of infinity, and by a null surface
which is spatially noncompact. We will call this null surface the ``acceleration horizon'' ${\cal{H}}_{acc}$. On a constant time surface, a black
hole horizon ${\cal{H}}_{bh}$ is compact with $S^{2}$ topology, whereas ${\cal{H}}_{acc}$ is noncompact with planar topology. Also, following the
common usage, we will often refer to the spatial sections of a horizon as just the ``{\it horizon}''. (For example, in the first law one talks about
the change in the area of the horizon of the black hole, meaning the area of a spatial cross section.) Hence we will refer to the black hole horizon
as compact, and the acceleration horizon as noncompact.

To summarize, a  boost vector comes with an acceleration horizon. The acceleration horizon is noncompact, and this raises finiteness questions in the
first law involving changes in the area of ${\cal{H}}_{acc}$, which we will address in section \ref{mboost}.

\subsection{$M_{boost}$ for Asymptotically Rindler Spacetime with a \\ Missing Angle}\label{mboost}

In this section we define the Boost mass for spacetimes that are asymptotically Rindler spacetime with a missing angle. The definitiion of $M_{Boost}$ only requires that spacetime have a boost Killing vector asymptotically.

Consider flat space minus a wedge spacetime with angular deficit parameter $\nu$
\begin{equation}
ds^{2} = -{dt^{\prime}}^{2} + {dz^{\prime}}^{2} + d\rho^{2} +
\nu^{2} \rho^{2}d\phi^{2}.\label{flatminus}
\end{equation}
For brevity, as in the introduction we will simply write this as $\eta^{\scriptscriptstyle{(\nu)}}_{ab}$ \footnote{However, for metrices which are not
everywhere equal to the refernce such as C or Ernst metric, the geometry specifies a value of $\kappa_{acc}$}. Let $z^{\prime} = z \hspace*{2pt}cosh
(\kappa_{acc}t)$ and $t^{\prime} = z\hspace*{2pt} sinh (\kappa_{acc}t)$. Then the previous equation becomes Rindler spacetime
 with a missing angle, given by the two parameter metric
\begin{equation}
ds^{2} = -\kappa_{acc}^{2}z^{2}dt^{2} + dz^{2} + d\rho^{2} + \nu^{2}
\rho^{2}d\phi^{2}.\label{rindler}
\end{equation}
The range of $\phi$ is $ 0 < \phi \leq 2\pi$. We will be concerned with the region $z> 0$.  Here $\kappa_{acc}$ is the surface gravity of the
acceleration horizon and $\nu$ is the angular deficit parameter around the $\phi$ axis. This metric describes the spacetime outside an infinite
straight static string and we will refer to it as a cosmic string.

$\vec T = \frac{\partial}{\partial t^{\prime}} = (\frac{1}{z \kappa_{acc}}cosh t \frac{\partial}{\partial t} - sinh t \frac{\partial}{\partial z}) $
is a Killing vector which translates in the $t^{\prime}$ time and is timelike everywhere. $\vec{\xi}= z^{\prime}\frac{\partial}{\partial t^{\prime}}+
t^{\prime}\frac{\partial}{\partial z^{\prime}}=\frac{1}{\kappa_{acc}} \frac{\partial}{\partial t}$ is another Killing vector, which translates in the
Rindler time $t$. It is only timelike in the region $S$, ${z^{\prime}}^{2}> {t^{\prime}}^{2}$ and becomes null on the surfaces ${z^{\prime}} = \pm
{t^{\prime}}$. $\xi^{a}$ is usually called a boost Killing vector.

We now consider a spacetime with metric $g_{ab}$ which approaches $\eta_{ab}^{\scriptscriptstyle{(\nu)}}$ as $\rho \rightarrow \infty$ in $z^{\prime}
\geq 0$ region, i.e, $g_{ab} = \eta_{ab}^{\scriptscriptstyle{(\nu)}} + ^{\scriptscriptstyle{(4)}}\gamma_{ab}$ and $\gamma_{ab}$ goes to zero asymptotically. Hence, $g_{ab}$ is not asymptotically flat, since the background spacetime contains the stress energy of an infinite cosmic string.

We define the boost charge of $g_{ab}$ by the general expression in equation (\ref{charge}), with $\vec{V}$ taken to be the asymptotic boost Killing vector of $g_{ab}$. The definition of $Q_{boost}(\vec{\xi})$ only depends upon the spacetime fields in the asymptotic region; $\vec{\xi}$ need not be a Killing field everywhere. This may seem to be at odds with the fact the acceleration horizon of a boost Killing vector extends into the interior of the spacetime. The point is that the background spacetime $\eta_{ab}^{\scriptscriptstyle{(\nu)}}$ does have a boost Killing field $\vec{\xi_{(0)}}$ everywhere, with an acceleration horizon. $\vec{\xi}$ is asymptotic to this background $\vec{\xi_{(0)}}$.

With $z = R cos\theta$ the metric in equation (\ref{rindler}) becomes
\begin{equation}
ds^{2} = -\kappa_{acc}^{2}R^{2} cos^{2}\theta dt^{2} + dR^{2} + R^{2}(d\theta^{2} +\nu^{2}sin^{2}\theta d\phi^{2}); ~~~~~ 0 < \theta \leq
\frac{\pi}{2}. \label{rindlerspherical}
\end{equation}

Let $\Sigma$ be a $t=$ constant spacelike slice which extends to spatial infinity. The unit normal vector to this surface is given by $\vec n
={(\kappa_{acc}R cos\theta)}^{-1}\frac{\partial}{\partial t}$. Now as $\vec{\xi} = F_{boost} \vec{n}$, the lapse function for the boost Killing field
is $F_{boost} = R cos\theta$ and the shift vector is zero. The area element on $\partial \Sigma$ is $da^{a} = \nu R^{2}sin\theta d\theta d\phi\
(\frac{\partial}{\partial R})^{a}$. In these coordinates the $S^{2}$ sphere has an area proportional to $R^{2}$. Following the general definition of
equation (\ref{charge}), the expression for the boost Killing charge with the background of flat space minus a wedge spacetime
 of equation (\ref{rindler}), is given by
\begin{equation}
16\pi Q_{boost} = -\int_{\partial \Sigma_{\infty}}da_{c}[F(D^{c}\gamma - D_{b}\gamma^{cb})-\gamma D^{c}F + \gamma^{c}_{b}D^{b}F], \label{boost}
\end{equation}
 Writing this out in terms of partial
derivatives we get
\begin{equation}
16\pi Q_{boost} = -\nu \lim_{R \to \infty} \int_{0}^{2\pi}d\phi \int_{0}^{\frac{\pi}{2}}d\theta[- \gamma^{\theta}_{R}\hspace*{2pt}R
\hspace*{2pt}sin\theta + R \hspace*{2pt}cos\theta \hspace*{2pt} \partial_{R}\gamma -\frac{1}{R}cot\theta \hspace*{2pt}\partial_{c}(R^{2}sin\theta
\gamma^{c}_{R})] R^{2}sin\theta,  \label{boostmass}
\end{equation}
where $\gamma=$Tr$[\gamma_{ab}]$. Instead of integrating over the whole $S^{2}$ sphere at spatial infinity, we integrate only over half of the
sphere, i.e $0 < \theta \leq \frac{\pi}{2}$, which bounds the region where the Killing field is timelike.

$Q_{boost}$ does not have the units of mass, rather it is dimensionless. This is because  we have written the boost Killing vector $\vec{\xi}$ so
that it is dimensionless, like a rotation, while the dimension of time translation  $[\vec{T}]=time^{-1}$. However since $Q_{boost}$ is associated
with time translation in Rindler time, it is nice to call it mass. To be consistent with units we define $M_{boost} = \kappa_{acc}Q_{boost}$. So
$M_{boost}$ is the charge associated with the rescaled vector $\vec{\xi} \rightarrow \kappa_{acc}\vec{\xi}$. Hereafter, we will refer to $M_{boost}$
(rather than $Q_{boost}$) which is more suited to comparison with $M_{ADM}$ and stating the first law in a familiar form, except for when analyzing
an example calculation of $\delta M_{boost}$ in section \ref{deltamboostforc}.

We now discuss the normalization of the boost vector $\vec{\xi}$, which involves the role of the parameter $\kappa_{acc}$ in equation (\ref{rindler}). Define the acceleration $a^{c}=\xi^{b}\nabla_{b}\xi^{c}$, and let $\vec{\xi}=N\frac{\partial}{\partial t}$. Then $\frac{|a^{c}a_{c}|}{|\xi^{b}\xi_{b}|}=\kappa_{acc}^{2}N^{2}$. We fix the normalization of $\vec{\xi}$, and hence the surface gravity of the acceleration horizon, by specifying the (dimensionful) parameter $\kappa_{acc}$, and requiring that $\frac{|a^{c}a_{c}|}{|\xi^{b}\xi_{b}|}=\kappa_{acc}^{2}$. For a particular metric describing an accelerating mass, the physics may pick out a preferred value of $\kappa_{acc}$. For example, a Rindler particle located at $z=\frac{1}{\kappa_{acc}}$ has four velocity $\vec{\xi}$, and constant acceleration $\vec{a}\cdot\vec{a}=\kappa_{acc}^{2}$.  Kinnersley and Walker \cite{C} have shown that the trajectories of the acceleration black holes in the C-metric are like those for constant acceleration Rindler particles, in the small mass limit.

A particular spacetime may have more than one symmetry. Depending on how a spacetime approaches the background, different charges may be zero, finite or infinite. In particular for a given
metric $g_{ab}$, not both of $M_{ADM}$ and $M_{boost}$ will be finite. This follows from the analysis of the boundary terms.

The ADM mass, results from the substitution of $\vec{T}$ in equation (\ref{charge}). On a constant $t$ slice the lapse function for the time
translation Killing field, $F_{TT} = cosh\hspace*{2pt}t$, is one power less in radial coordinates in comparison to the boost Killing field. Let
$\gamma_{\hat{i}}^{\hat{j}}$ be the components  of $\gamma_{ab}$ in an ortho-normal frame. Therefore, whereas for the $M_{ADM}$ to be finite we need
$\gamma_{\hat{i}}^{\hat{j}}$ to fall-off as $R^{-1}$, for the Boost charge to be finite we need $\gamma_{\hat{i}}^{\hat{j}}$ to fall off as $R^{-2}$.
This is evident from the expressions of equation (\ref{adm}) and equation (\ref{boostmass}). Hence for a given $\gamma^{j}_{i}$, $M_{ADM}$ and
$M_{boost}$ will not both be finite and nonzero. If the perturbations fall off as $R^{-1}$, $M_{ADM}$ is finite but $M_{boost}$ is infinite; if they
fall off as $R^{-2}$ the $M_{ADM}$ is zero but $M_{boost}$ is finite. The same analysis of the fall off conditions  for  the finiteness of $M_{ADM}$
and $M_{boost}$ will also apply for the perturbative charges $\delta M_{ADM}$ and $\delta M_{boost}$. We will see that the C and Ernst spacetimes
naturally yield examples with $ M_{boost} \neq 0$, but $ M_{ADM} =0$. More generally, Bicak and Schmidt \cite{bicak} have shown that besides the C
metric, there are boost and rotation symmetric spacetimes corresponding to general sources moving on boost-symmetric orbits.

\subsection{First Law for $\delta M_{boost}$} \label{Firstlaw}

We now derive the first law of black hole mechanics for $\delta M_{boost}$. Consider a background spacetime $g_{ab}^{\scriptscriptstyle(0)}$
which is asymptotic to the Rindler metric with a missing angle, given in equation (\ref{rindler}). We consider $\nu$, which determines the mass per
unit length of the string and $\kappa_{acc}$, the surface gravity  of the acceleration horizon, to be fixed at infinity. We assume that the metric
$g_{ab}^{\scriptscriptstyle(0)}$ has a boost Killing vector $\vec{\xi} = \frac{\partial }{{\partial t}}$ throughout the spacetime, with an associated acceleration horizon
${\cal{H}}_{acc}$, on which the Killing field $\xi^{a}$ goes null. In addition suppose that $g_{ab}^{\scriptscriptstyle(0)}$ has a black hole, and
that the black hole horizon is also generated by $\xi^{a}$. Since the boost Killing vector generates both the horizons, the resulting first law will
relate the variations of the areas of the different horizons to the variation of $M_{boost}$, the Killing charge corresponding to $\xi^{a}$, instead
of the usual ADM mass.

Now, we will study the perturbations about this background spacetime $g_{ab}^{\scriptscriptstyle(0)}$ i.e, $g_{ab}=g_{ab}^{\scriptscriptstyle(0)} + \
^{\scriptscriptstyle{(4)}}h_{ab}$ where the perturbations $ \ ^{\scriptscriptstyle{(4)}}h_{ab}$ satisfy the linearized Einstein equations, and hence
the linearized constraints. Consider a spacelike slice $\Sigma$ which intersects the black hole horizon on the bifurcation sphere and the acceleration horizon on the bifurcation surface. Substituting the Killing field $\xi^{a}$ in the constraint equation
(\ref{constraint}), the gravitational boundary terms at spatial infinity gives us $\delta M_{boost}$. The expression for $\delta M_{boost}$ is the
same as the expression for $M_{boost}$ given in equation equation (\ref{boostmass}), with $\gamma_{ab}$ replaced by $h_{ab}$. That is
\begin{equation}
16\pi \delta M_{boost} = -\nu \lim_{R \to \infty} \int_{0}^{2\pi}d\phi \int_{0}^{\frac{\pi}{2}}d\theta[- h^{\theta}_{R}\hspace*{2pt}R
\hspace*{2pt}sin\theta + R \hspace*{2pt}cos\theta \hspace*{2pt} \partial_{R}h -\frac{1}{R}cot\theta \hspace*{2pt}\partial_{c}(R^{2}sin\theta
h^{c}_{R})] R^{2}sin\theta.  \label{deltamboost}
\end{equation}

There are differences in the proof of the first law for $\delta M_{ADM}$ and $\delta M_{boost}$. These differences are important when the differential Gauss' Law of equation (\ref{GaussLawDiff}) is converted to the integral form of equation (\ref{constraint}). In the proof of the  first Law for a black hole in an asymptotically flat spacetime with a time translation Killing vector, there is a boundary at spatial infinity, and a boundary at the black hole horizon. With a boost Killing vector, in addition to the compact black hole horizon we also have a noncompact acceleration horizon, which extends to spatial infinity. Further, the spacetime is not flat at infinity.  Since there is an infinite amount of string at infinity, even $\delta M_{boost}$ would be infinite if we were comparing to Minkowski spacetime. However, the definition of $\delta M_{boost}$ in equation (\ref{deltamboost}) compares the perturbed spacetime to the background of a cosmic string. The result may still be infinite, but it may also be finite.

There are two additional boundaries in this case, one at each horizon. $\xi^{a}$ vanishes on the bifurcation surface of either horizon. The
derivative of the lapse function $F$ along the normal to the bifurcation surface is proportional to the surface gravity $\kappa$ of the respective
horizon. Since $\kappa$ is constant over the bifuraction surface, this can be taken out of the integral and the gravitational boundary term on each
horizon reduces to
\begin{equation}
\int_{\cal{H}} da_{c}(-hD^{c}F + h^{cb}D_{b}F) = \kappa \int_{\cal{H}}(h^{x^{1}}_{x^{1}}+ h^{x^{2}}_{x^{2}})da = 2\kappa \delta A.
\label{deltaaacceleration}
\end{equation}
Here $x^{1}$ and $x^{2}$ are coordinates on the horizon and $da$ is the area element of the spatial metric on the horizon.

It is worthwhile to first consider the case when no black holes are present, and there is no charge. This allows us to focus on  the new features due
to the boost Killing vector. Using equations (\ref{deltamboost}) and (\ref{deltaaacceleration}) in equation (\ref{constraint}) it gives the first law for
acceleration horizons,
\begin{equation}
\delta M_{boost}=
\frac{1}{8\pi}\kappa_{Acc}\delta A_{acc} + \int_{\Sigma} \xi^{a}n_b\delta T_a ^b .
\label{firstlawacc}
\end{equation}

The first issue is to determine when the  various terms in equation (\ref{firstlawacc}) are finite. The conditions for finiteness of $\delta M_{boost}$
are completely analogous to those for $M_{boost}$, namely that $\delta M_{boost}$ is finite when  $h^{\hat{i}}_{\hat{j}} \rightarrow \frac
{1}{R^{2}}$ at spatial infinity, where the coordinate $R$ is defined in equation (\ref{rindlerspherical}). Next, since the acceleration horizon
itself is noncompact, finiteness of $\delta A_{acc}$ is an issue. Suppose there are no perturbative sources, $i.e., \  \delta T^a _b =0$. Equation
(\ref{firstlawacc}) is an identity on the solutions to the linearized equations (about $g_{ab}^{\scriptscriptstyle(0)}$). Therefore, if $\delta
M_{boost}$ is finite, $\delta A_{acc}$ must also be finite. From this point of view, a divergent $\delta A_{acc}$ is simply associated with a
divergent $\delta M_{boost}$, which diverges {\it more readily } than $\delta M_{ADM}$ because the boost Killing vector diverges at infinity.

Examples of finite changes $\delta A_{acc}$ have been calculated in particular source free cases, using the C-metric and Ernst spacetime, in
references \cite{breaking cosmic string} and \cite{Hawking:1994ii}, respectively. In the next section, we will compute $\delta M_{boost}$ in an example involving C-metric.

Now consider perturbative sources $\delta T_a ^b$. If the sources do not have compact support then fall-off conditions are necessary so that the
volume integral of $ \xi^{a}n_{b}\delta T_a ^b$ is finite. For asymptotically flat spacetimes with a time translation Killing vector, the source
integral goes like the monopole moment of $\delta\rho$. By contrast, with the boost Killing field, the source integral goes like $z\delta\rho$, at
large $z$, which is a dipole moment of the source. So finiteness of $\delta M_{boost}$ requires that the dipole moment of the source is finite,
whereas finiteness of $\delta M_{ADM}$ only requires a finite monopole moment. This is just the same condition that we have already seen for the
respective rates of fall off of the metric perturbation $h_{ab}$ in the far field. The two are connected since $h_{ab}$ is the solution to a
Poisson-type equation with source $\int_{\Sigma} \xi^{a}n_{b}\delta T_a ^b$.

We also gain some understanding of the physical meaning of  $\delta M_{ADM}$ and  $\delta M_{boost}$ by focusing on the relation between these mass
variations and the matter sources. In the  weakly gravitating, but still relativistic limit, we have from equations (\ref{firstlawadm}) and
(\ref{firstlawacc})
\begin{equation}
\delta M_{ADM} \sim \int dv \delta T_{\hat t  \hat t }
\end{equation}
whereas,
\begin{equation}
\delta M_{boost}\sim
\frac{1}{8\pi}\kappa_{Acc}\delta A_{acc}  +\int dv ( z\delta T_{\hat t \hat t } - t \delta T_{\hat t \hat z } ).
\end{equation}
Here the hats denote an ortho-normal frame. These relations help justify the names of the charges. The first is the Newtonian relation that follows
from Poisson's equation, if one judiciously defines the total mass of the system by a boundary integral of the gradient of the Newtonian potential.
The source of the ``Noether time-translation charge'' is the mass density! The second relation says that the source of the ``Noether boost-invariance
charge'' is the boost-moment of the stress-energy. (On a $t=0$ surface this becomes a dipole moment of the energy density.) Because the boost mass is
generated by a boost Killing vector, the $\delta A_{acc}$ term is still present. It would be interesting to know if there are any solutions to the
Einstein's equation in which $\delta A_{acc}$ is zero, and $\delta M_{boost}$ is due  just to the source terms.

Having already considered the issues of convergence, adding the black hole to the first law for boost mass is simple. Again, using
equation(\ref{constraint}) with the additional internal black hole horizon boundary, we have the first law,
\begin{equation}
\delta M_{boost}=\frac{1}{8\pi}\kappa _{BH}\delta A_{BH} +
\frac{1}{8\pi}\kappa_{Acc}\delta A_{acc} - <A_{t}\delta Q> +
\int_{\Sigma} \frac{(\xi^{a}n_{b}\delta T_{a}^{b})}{16\pi}.\label{firstlaw}
\end{equation}
where
\begin{equation}
<A_{t}\delta Q>=\int_{\partial \Sigma_{\infty}}da_{c}\frac{1}{\sqrt{s}}A_{t}\delta p^{c},
\end{equation}
and $Q$ is the electric charge, $\delta Q = \int_{\partial \Sigma_{\infty}}\frac{1}{\sqrt{s}}da_{c}\delta p^{c}$. We have choosen the gauge potential
such a way that it vanishes on the horizons. When $A_{t}$ is constant on the boundary at infinity then $<A_{t}\delta Q> = A_{t}\delta Q$. Equation
(\ref{firstlaw}) is the desired first Law. It holds for any solution to linearized Einstein's equation, when the background spacetime has a boost
Killing vector with bifurcate black hole and acceleration horizon.

Based on the local notion of horizon entropy density, Jacobson and Parentani \cite{Jacobson} argue that the laws of black hole thermodynamics apply quite
generally to any causal horizon. That work discusses the first Law including the change in the area  of acceleration horizon (there called Rindler
Horizon), in asymptotically flat spacetime where the background metric is Minkowski. We have seen that in this case the the boost mass of each spacetime is infinite. So the Minkowski background is not an intersting case. For example, if
one compares the spacetime of an isolated mass to Minkowski spacetime, the boost mass of the spacetime is infinite (and Minkowski spacetime has
$M_{boost} = 0$). Since $\delta M_{boost}$ is infinite, $\delta A_{acc}$ is also infinite. This is independent of the type of motion of that mass.

Note that it is possible that the difference $\delta M_{boost} = M_{boost}^{\scriptscriptstyle{(2)}} - M_{boost}^{\scriptscriptstyle{(1)}}$ between two boost masses is finite for two
asymptotically flat metrices $g_{ab}^{\scriptscriptstyle{(2)}}$ and $g_{ab}^{\scriptscriptstyle{(1)}}$. This just requires $g^{\scriptscriptstyle{(1)}}_{ab}$ and $g^{\scriptscriptstyle{(2)}}_{ab}$ have the same ADM masses.
However, in this case we do not have the first law of equation (\ref{firstlaw}) for $\delta M_{boost}$. This is because neither $g^{\scriptscriptstyle{(1)}}_{ab}$ or
$g^{\scriptscriptstyle{(2)}}_{ab}$ will have a boost Killing vector. If the spacetime is asymptotically flat but contains stress energy, it will not have a boost
symmetry.

\section{An Example with the C-metric}\label{cst}

\subsection{The C-metric}

An example of interest, indeed the motivation for this work, is to choose the background spacetime to be a C-metric \cite{C}. The C-metric describes
the spacetime corresponding to two charged black holes of opposite charge, uniformly accelerating away from each other along a symmetry
axis, being pulled apart by a cosmic string. More precisely, the electrically charged C-metric is given by
\begin{equation}
ds^2 = \frac{1}{A^2(x - y)^2}[{G(y)dt^2 - G^{-1}(y)dy^2 +
G^{-1}(x)dx^2 +\mu^{2}G(x)d{\phi}^2}],
\label{chargedc}
\end{equation}
\[
 A_{t} = \sqrt{r_{+}r_{-}}y \hspace*{3pt},
\hspace*{20pt} G(\xi) = (1 +r_{-}A\xi)[1 - {\xi}^2(1 +
r_{+}A\xi)].
\]
Here $\phi$ ranges from $0$ to $2\pi$.

The metric has two Killing vectors, $\frac{\partial}{\partial t}$ and $\frac{\partial}{\partial \phi}$. The horizons occur where norm of
$\frac{\partial}{\partial t}$ vanishes i.e., at the zeroes of $G(y)$. Let $\xi_{1 }\equiv -\frac{1}{r_{-}A}, \xi_{2}, \xi_{3}$, and $\xi_{4}$ be the
four real roots of $G(\xi)$, which exist for $ r_ +  A < 2/3\sqrt 3$. The surface $y = \xi_{2}$ is the compact black hole horizon and the surface $y
= \xi_{3}$ is the noncompact acceleration horizon; they both are Killing horizons for $\frac{\partial}{\partial t}$. For the range of coordinates
$\xi_{2} \leq y \leq \xi_{3}$, $G(y)$ is negative and hence the Killing field $\frac{\partial}{\partial t}$ is time like. Thus the Killing field
$\frac{\partial}{\partial t}$ is timelike in a part of the spacetime which is bounded by the acceleration horizon, black hole horizon and a part of
spatial infinity. Therefore, according to our definition in section \ref{definitions} we identify the Killing field $\frac{\partial}{\partial t}$
as a boost Killing field.

The coordinates $(x,\phi)$ are angular coordinates. $\xi_{3} < x < \xi_{4},$ where $G(x)$ is positive. The norm of the Killing vector
$\frac{\partial}{\partial \phi}$ vanishes on the axis, at $x = \xi_{3}$ and $ x= \xi_{4}$. The axis $x = \xi_{3}$ extends to spatial infinity. The
axis $x = \xi_{4}$ points towards the other black hole. Spatial infinity is reached by fixing $t$ and letting both $y $ and $x$ approach $\xi_{3}.$

The C metric has conical singularities on the symmetry axis. The deficit angle $\delta_{in} =2\pi[1-(\frac{\mu}{2})|G^{\prime}(\xi_{4})|]$ is on the
inner part of the axis which is between the two black holes. On the outer part of the axis, extending from each black hole to infinity, the deficit
angle is $\delta_{out} = 2\pi[1-(\frac{\mu}{2})|G^{\prime}(\xi_{3})|]$. We interpret the conical singularities as a model for a thin cosmic string
along the symmetry axis. We take $\delta_{in} < \delta_{out}$, $i.e.,$ the mass per unit length of the string is greater on the outer axis than on
the inner axis. The corresponding difference in string tension between the outer and inner part of the axis provides the force which accelerates the
black holes.

The C-metric has four parameters $r_{+},r_{-}, A, \mu$. Define $m$ and $q$ via $ m = \frac{1}{2}(r_{+} + r_{-}), \hspace*{1pt} q =\sqrt{
r_{+}r_{-}}$. Then in the limit $r_{\pm}A \ll 1$, $m, q, A$ and $\mu$ denote the mass, charge, acceleration of the black holes and the mass per unit
length of the cosmic string respectively \cite{dggh, C}.

At spatial infinity, the C-metric is asymptotic to Rindler spacetime minus a wedge. To see this we consider a particular C-metric and expand the
function $G(\xi)$ around the spatial infinity, $x,y \to \xi_{3}$, i.e $G(\xi) = G(\xi_{3}) + \lambda (\xi - \xi_{3})+\beta(\xi - \xi_{3})^2$, where
$\lambda = \left. {\frac{{dG}}{{d\xi }}} \right|_{\xi  = \xi _3 }$ and $\beta=\left.\frac{1}{2}{\frac{{d^2G}}{{d\xi^2 }}} \right|_{\xi  = \xi _3 }$.
Make the transformations $(y-\xi_{3}) = -\frac{4}{A^2\lambda}\frac{cos^{2}\alpha}{\tilde{r}^{2}}$ and $(x- \xi_{3}) =
\frac{4}{A^2\lambda}\frac{sin^{2}\alpha}{\tilde{r}^{2}}$ where $ 0 < \alpha \leq \frac{\pi}{2} $. Rescaling the time coordinate so that it has the
dimensions of time $\tilde{t} = \frac{t}{A}$, we get the asymptotic form of the metric at spatial infinity $\tilde{r}\rightarrow \infty $ as
\begin{eqnarray}
ds^{2} & \to & -\kappa_{acc}^{2}\tilde{r}^{2}cos^{2}\alpha(1 - \frac{\beta}{{\kappa^{2}_{acc}}\tilde{r}^{2}}cos^{2}\alpha) d\tilde{t}^{2} +(1+
\frac{\beta}{{\kappa^{2}_{acc}}\tilde{r}^{2}}cos2\alpha) d\tilde{r}^{2} + \tilde{r}^{2}d\alpha^{2} \nonumber \\ & & + \frac{\beta}{2 \kappa^{2}_{acc}\tilde{r}}sin 2\alpha d\tilde{r}d\alpha+
\tilde{r}^{2}\nu^{2}sin^{2}\alpha ( 1+ \frac{\beta}{{\kappa^{2}_{acc}}\tilde{r}^{2}}sin^{2}\alpha) d\phi^{2}.
\label{asymcmetric}
\end{eqnarray}
Here $\kappa_{acc} = \frac{1}{2}A\lambda$ and $\nu = \frac{1}{2}\mu\lambda$.

For $r_{\pm}=0$, then $\xi_{3} = -1$,  $\lambda = 2$, and $\beta = -1$. Hence equation(\ref{asymcmetric}) becomes
\begin{eqnarray}
ds^{2} & \to & -\kappa_{acc}^{2}\tilde{r}^{2}cos^{2}\alpha(1 + \frac{1}{{\kappa^{2}_{acc}}\tilde{r}^{2}}cos^{2}\alpha) d\tilde{t}^{2} +(1-
\frac{1}{{\kappa^{2}_{acc}}\tilde{r}^{2}}cos2\alpha) d\tilde{r}^{2} + \tilde{r}^{2}d\alpha^{2} \nonumber \\ & & -
\frac{1}{2 \kappa^{2}_{acc}\tilde{r}}sin 2\alpha d\tilde{r}d\alpha +
\tilde{r}^{2}\nu^{2}sin^{2}\alpha ( 1- \frac{1}{{\kappa^{2}_{acc}}\tilde{r}^{2}}sin^{2}\alpha) d\phi^{2}.
\label{refst}
\end{eqnarray}
where $\kappa_{acc} = A$ and $\nu= \mu$. Equation (\ref{refst}) is our reference spacetime, but it is not of the same form as Rindler equation
(\ref{rindlerspherical}). This just means that the $(R,\theta)$ coordinates of equation (\ref{rindlerspherical}) are not the same as the
$(\tilde{r},\alpha)$ coordinates of equation (\ref{refst}). For $r_{\pm}=0$, Kinnersley and Walker \cite{C} have given a sequence of coordinate
transformations that take the C-metric to flat spacetime minus a wedge of equation(\ref{rindler}). However, it is nontrivial to work in their
coordinates when $r_{\pm} \neq 0$. To calculate $\delta M_{boost}$ we will need equation (\ref{asymcmetric}), in which $\beta$ is general.

To summarize, equation (\ref{refst}) is Rindler spacetime with a missing angle $\eta^{\scriptscriptstyle{(\nu)}}_{ab}$, which is our reference
metric. The asymptotic form of a general C-metric is given by equation (\ref{asymcmetric}). The reference spacetime is fixed by specifying
$\kappa_{acc}$ and $\nu$. Physically this means fixing surface gravity of the acceleration horizon and mass per unit length of the cosmic string,
both at infinity. The initial C-metric has four parameters, $r_{+},r_{-},A,\mu$ and when we perturb this metric to another close by C-metric (section
\ref{deltamboostforc}), $\kappa_{acc}$ and $\nu$ are kept fixed. This leaves a 2-parameter family of solutions to the linearzied equtions.

\subsection{$M_{boost}$ and $M_{ADM}$ for C-metric}

Let $g_{ab}^{\scriptscriptstyle{(0)}}$ be a C-metric with particular values of $\kappa_{acc}$ and $\nu$ as defined in equation (\ref{rindler}). The
C-metric has both time translation and boost symmetry asymptotically, though only the boost is a Killing vector throughout the spacetime. Using these
asymptotic Killing vectors, we can compute $M_{boost}$ [see equation (\ref{boostmass})] and $M_{ADM}$ [see equation (\ref{adm}), with $h_{ab}$
replaced by $\gamma_{ab}$]. To do this, we need the far field $\gamma_{ab}$ of the metric near spatial infinity, where $\gamma_{ab} =
g_{ab}^{\scriptscriptstyle{(0)}} - \eta_{ab}^{\scriptscriptstyle{(\nu)}}$. The components in an ortho-normal frame are
\begin{eqnarray}
\gamma^{\hat t}_{\hat t} = -\frac{\Delta\beta}{\kappa_{acc}^{2}{\tilde r}^2}cos^{2}\alpha;\hspace*{16pt} \gamma^{\hat r}_{\hat r} =
\frac{\Delta\beta}{\kappa_{acc}^{2}{\tilde r}^2}cos2\alpha;\hspace*{16pt}
\gamma^{\hat \alpha}_{\hat \alpha} = 0; \nonumber\\
\hspace*{16pt} \gamma^{\hat \phi}_{\hat \phi}= \frac{\Delta\beta}{\kappa_{acc}^{2}{\tilde r}^2}sin^{2}\alpha;\hspace*{16pt} \gamma^{\hat r}_{\hat
\alpha}= \frac{\Delta\beta}{2\kappa_{acc}^{2}{\tilde r}^{2}}sin2\alpha;\hspace*{16pt}\gamma^{\hat \alpha}_{\hat r}=
\frac{\Delta\beta}{2\kappa_{acc}^{2}{\tilde r}^2}sin2\alpha, \label{perturbations}
\end{eqnarray}
where $\Delta \beta = \beta_{\scriptscriptstyle{(0)}}-(-1)$. Note that all the perturbations are propotional to $\Delta \beta$.

Now, we can compute $M_{boost}$ with $r_{\pm} \neq 0$. On a constant time slice the lapse function is $F= - \hat{n}\cdot \vec{\xi} = \tilde{r}
cos\alpha$ and the area element $da^{a} \sim \tilde{r}^{2}$. Plugging the perturbations in equation (\ref{boostmass}), we get $M_{boost} =
(\frac{\nu}{8\kappa_{acc}})\Delta \beta$, a finite and nonzero result. In the next section we will solve for $\Delta \beta$ in terms of $m$ and $q$
for $Am, Aq << 1$.

In a similar way we can compute $M_{ADM}$. For simplicity, evaluate the integral on a constant time $t=0$ slice. Then $F=- \vec{T}.\vec{n} = 1$. The
perturbations fall off as $\gamma^{\hat{i}}_{\hat{j}} \sim \frac{1}{\tilde{r}^{2}}$, and therefore following the general definition of gravitational
charge, we find $M_{ADM} =0$. This is simply because fixing $\kappa_{acc}$ and $\nu$ at infinity is equivalent to fixing monopole moment of the
system.  The C-metric is a sort of {\it dipole rearranging} of the reference spacetime - some of the string mass goes into black hole mass or vice
versa. To summarize, the charge of interest for the C-metric is the boost mass, which measures a dipole rearrangement of the background
stress-energy.

\subsection{$\delta M_{boost}$ for the C-metric} \label{deltamboostforc}

In this section we will compare two nearby C-metrics to find the perturbative charge defined in equation (\ref{perturbativecharge}). More
specifically, we calculate $\delta M_{boost}$ between two C-metric spacetimes. Take a particular C-metric as the background spacetime
$g^{\scriptscriptstyle{(0)}}_{ab}$. Fixing $\nu$ and $\kappa_{acc}$, choose another nearby C-metric $g_{ab}$. Let $^{\scriptscriptstyle{(4)}}h_{ab} =
g_{ab}-g^{\scriptscriptstyle{(0)}}_{ab}$ and $\delta A_{b}$ be the perturbations to linear order in $\delta A,\delta {\mu}, \delta r_{\pm}$. So
$(^{\scriptscriptstyle{(4)}}h_{ab}, \delta A_{b})$ is a solution to the linearized Einstein equation, with no sources. Note that, at infinity ${h_{ab}}$ is same as
${\gamma_{ab}}$ with $\Delta \beta \equiv \beta - (-1)$ replaced by $\delta \beta \equiv \beta [g] - \beta[g^{\scriptscriptstyle{(0)}}]$. Thus
$\delta M_{boost} = (\frac{\nu}{8\kappa_{acc}})\delta \beta$.

Next we rewrite this result in a more meaningful way. Take  the $g_{ab}^{\scriptscriptstyle{(0)}}$ to be just  a string, with no black holes, $
g_{ab} ^{\scriptscriptstyle{(0)}} =\eta _{ab}^{\scriptscriptstyle{(\nu)}}$ and let $g_{ab}$ be a C-metric with small black
holes and the same $\kappa _{acc} ,\nu$. The expressions  can be simplified for $Ar_{\pm}\ll 1$. To leading order we find that $\nu ={
\mu}(1-2Am)$ and $\kappa _{acc} =A(1-2Am)$. The missing angle parameter on the inner axis is given by $\nu _{in} ={\mu} (1+2Am)$, which is
always greater than $\nu$, $i.e.$, the deficit angle on the inner axis is less than the deficit angle on the outer axis. This also means that the
metric parameter ${\mu}$ is constrained by ${\mu} \leq (1-2Am)$. Lastly, $\delta\beta =6Am$ and therefore
\begin{equation}\label{deltamforc}
\delta M_{boost} ={3\over 4}\nu  m.
\end{equation}
This result is proportional to the black hole mass parameter times the angle deficit parameter. So for fixed $m$, the boost mass decreases as the
outer deficit angle increases, using $\nu = (1-{\delta _{out} \over 2\pi })$. The proportionality to $\nu$ is due to the modification in the area of
two-spheres from the missing angle. One can see this by computing the ADM mass for Schwarzchild with a missing angle. In the usual Schwarzchild
metric, replace $d\phi ^2$ by $\nu ^2 d\phi ^2$. This is still a vacuum solution to the Einstein equation. Then the only change in the integral for
the ADM mass is that the area element has a factor of $\nu$, just as in computing $\delta M_{boost}$.

Rewriting equation (\ref{deltamforc}) in terms of the boost charge ($\delta Q_{boost} ={1\over \kappa_{acc}} \delta M_{boost}$ and $\xi^{a}$ is
dimensionless) highlights the difference between the boost charge and the ADM mass. We have
\begin{equation}\label{deltaq}
\delta Q_{boost} ={1\over \kappa_{acc}} \delta M_{boost} = {1\over \kappa_{acc}}{3\over 4}\nu  m.
\end{equation}
Suppose we were computing the boost mass for a Rindler particle of mass $m$,  instead of a black hole. The Rindler particle moves on the hyperbola
$-(t')^2 +(z')^2 = \kappa _{acc}^{-2}$ in the coordinates of equation (\ref{flatminus}), for which $t'$ is the particle's proper time. The surface
$t' =0$ coincides with the surface of constant Rindler time $t=0$, and we can approximate the source integral  for $\delta Q_{boost}$ on that slice:
$\int dv \delta T^a _b \xi^b n_a =\int dv z' \delta\rho \sim \kappa_{acc}^{-1}\nu m $. That is, $\delta Q_{boost}$ is the dipole moment of the
Rindler particle, where the  length of  the moment arm  $\kappa_{acc}^{-1}$ is the semi-major axis of the hyperbola.

Note that although it is tempting to compare this result to the analogous result for Schwarzchild, one cannot take the limit where the background
goes to flat space in equation (\ref{deltamforc}). Taking both $\nu \rightarrow 1$ and $\nu _{in}\rightarrow 1$ requires that $Am=0$. Further, there
is no reason to expect that the answers would be the same, since the Killing vectors are different vector fields.

It would be nice to check these results by computing $\delta Q_{boost}$, $\delta A_{bh}$, and $\delta A_{acc}$, and substituting in the first law, equation (\ref{firstlaw}). However, in the coordinates of equation (\ref{chargedc}) the C-metric is badly behaved on the horizons, and one finds that in particular the metric perturbation $h_{ab}$ is badly behaved. One would need to first find better, Kruskal-type coordinates near the horizons, and then expand to find $h_{ab}$. We leave this excercise to future work.

We close this section with some comments about the variation of the ADM mass for perturbations about a C-metric. Since the metric is asymptotically
Rindler spacetime with a time translation Killing vector $T^a$, the ADM mass is defined. However, since $T^a$ is not a Killing vector throughout the
spacetime, we do not have a theorem which relates $\delta M_{ADM}$ to the variations in the horizon areas. Still, one can compute $\delta M_{ADM}$.
If we compare the ADM mass for two perturbatively close  C-metrics with the same value of $\kappa_{acc}$ and $\nu$ then $\delta M_{ADM} = 0$. The
perturbations are given in (\ref{perturbations}), and since $F=-T^a n_a$ goes to a constant, the boundary term vanishes. On the other hand, if two
C-metrics are compared that have different values of  $\nu$, then $\delta M_{ADM} = \infty$. This is essentially because of the fact that  a change
in the mass per unit length over an infinite length is infinite. By contrast, if one adds a small mass source to the C-metric--say a planet orbiting
a black hole--then one expects the change in the ADM mass to be finite, and the change in the boost mass to be infinite.

\section{The Ernst Spacetime and Conclusions}\label{Ernstst}

We have seen that the boost mass is a relevant charge for a spacetime which has stress energy at infinity, and of course, has an asymptotic boost
Killing vector. If a metric with black holes has an exact boost symmetry, then we have shown that perturbations about the metric obey the first law
of black hole mechanics. This work was motivated  by studying the C-metric and Ernst metric, so we briefly mention the latter.

The Ernst spacetime \cite{Ernst} is another analytic solution to the Einstein-Maxwell theory, which  has a boost Killing vector. This spacetime
represents two oppositely charged black holes, undergoing uniform acceleration by a background magnetic field.

The Ernst metric has two Killing vectors: $\frac{\partial}{\partial t}$ and $\frac{\partial}{\partial \phi}$. Killing field $\frac{\partial}{\partial
t}$ becomes null on the compact black hole horizon, and on the noncompact acceleration horizon. The Killing field $\frac{\partial}{\partial t}$ is
timelike in a region of the spacetime which is bounded by the black hole horizon, the acceleration horizon and a part of spatial infinity. Thus
according to the definition in section \ref{definitions}, $\frac{\partial}{\partial t}$ is a boost Killing vector. At large spatial distances the
Ernst metric reduces to the Melvin metric \cite{melvin}, which contains a magnetic field throughout the spacetime. We consider Melvin sapcetime as
our reference spacetime $g_{ab}^{\scriptscriptstyle{(ref)}}$. Since the reference spacetime has nonzero stress energy in the asymptotic region, the
Ernst metric is not asymptotically flat. Following the general definition of equation (\ref{charge}) we can define the charge $M_{boost}$ for the
Ernst spacetime, corresponding to the asymptotic boost Killing field of Melvin spacetime. Similar to the C-metric, the Ernst spacetime has a finite,
nonzero boost mass. In addition to the black hole horizon, the Ernst metric has a spatially noncompact acceleration horizon both of which are
generated by $\frac{\partial}{\partial t}$.

Now consider a background black hole spacetime spacetime $g_{ab}^{\scriptscriptstyle{(0)}}$ that has a boost Killing field, and which is asymptotic
to Melvin, such as the Ernst metric. Fix the value of the surface gravity of the acceleration horizon  and  the magnetic field at spatial infinity.
Following the general derivation in section \ref{Firstlaw}, we can prove the first law  of equation (\ref{firstlaw}) for perturbations about
accelerated black holes in an asymptotically Melvin Universe.

Of course, there are many open issues.
It would be interesting to see if an analogous first law holds for higher dimensional black objects,
for example, black strings being pulled apart by two-branes,
 or charged strings being accelerated apart by
an external field. Another situation of interest would be to study the boost mass constraints
in the context of brane world scenarios, with a boost symmetric background brane. Here one appreciates the importance of which boundary conditions are appropriate. In a brane-cosmology,
depending on how cosmological perturbations are generated / present in initial conditions,
$\delta M_{boost}$ could be infinite, finite, or zero.

One would like to see if the accelerated black hole mechanics discussed here is
actually part of a thermodynamic structure. Three further key elements are needed; first whether, or not,
there is an area increase theorem for black holes in asymptotically cosmic string / Melvin spacetimes.
Second, if there is an area increase theorem for acceleration horizons in these cases.
And third, a calculation of Hawing radiation in these spacetimes.

Lastly, it would be interesting to have a definition of acceleration horizon and accelerating
black holes in the absence of a boost Killing vector. It may be that the best definition
of  black holes with constant acceleration is that the generator of the black hole horizon is a boost
Killing vector of the spacetime. What about the case of non-constant acceleration?
 In a test particle limit, one can talk about the acceleration of the
particles, and if these are black holes instead of particles, presumably one can talk about
the acceleration of the the black holes. What is the description if the black holes have
significant mass, but there is no spacetime symmetry? Is there an acceleration horizon in
some meaningful sense, or is this notion special to the case of a boost Killing vector?

\bigskip
\noindent {\bf Acknowledgements:}
We would like to thank David Kastor for useful conversations and for commenting on the draft. This work was
supported in part by U.S National Science Foundation Grant No. NSF PHY-02-44081.

\end{document}